\documentclass{aa}  

\usepackage[varg]{txfonts}

\usepackage{natbib}
\usepackage{graphicx}
\usepackage{dcolumn}
\usepackage{bm}
\usepackage{amssymb,amsmath,bm}
\usepackage{color}
\usepackage[colorlinks,linkcolor=red,citecolor=blue,urlcolor=blue ]{hyperref}
\usepackage{multirow}
\usepackage[utf8]{inputenc}
\usepackage{enumitem}

\makeatletter
\renewcommand*\aa@pageof{, page \thepage{} of \pageref*{LastPage}}

\newcommand{\rd}{\mathrm{d}}

\graphicspath{%
	{./figures/} %
}	

\begin{document}

\title{Cosmology with the kinetic Sunyaev--Zeldovich effect: Independent of the optical depth and $\sigma_8$}
\titlerunning{Cosmology with the kSZ}

\author{Joseph Kuruvilla}

\offprints{joseph.kuruvilla@universite-paris-saclay.fr}

\institute{Universit\'e Paris-Saclay, CNRS,  Institut d'Astrophysique Spatiale, 91405, Orsay, France}

\date{Received 2021; accepted}
\abstract{
\noindent The cosmological constraints from the kinetic Sunyaev--Zeldovich experiments are degenerate with the optical depth measurement, which is commonly known as the optical-depth degeneracy. In this work, we introduce a new statistic based on the first moment of relative velocity between pairs in a triplet, which is capable of constraining cosmological parameters independent of the optical depth, and $\sigma_8$. Using 22,000 $N$-body simulations from the Quijote suite, we quantify the information content in the new statistic using Fisher matrix forecast. We find that it is able to obtain strong constraints on the cosmological parameters, particularly on the summed neutrino mass. The constraints have a factor of 6.2--12.9, and 2.3--5.7 improvement on all cosmological model parameters when compared to those obtained from the mean pairwise velocity, and the redshift-space halo power spectrum, respectively. Thus the new statistic paves a way forward to constrain cosmological parameters independent of the optical depth and $\sigma_8$ using data from future kinetic Sunyaev--Zeldovich experiments alone.}


\defcitealias{Planck16}{Planck~Collaboration~XXXVII~2016}
\defcitealias{Planck2018}{Planck~Collaboration~VI~2020}
\defcitealias{eBOSS20}{eBOSS~Collaboration~2021}

\maketitle
\section{Introduction}

In modern physics, one of the outstanding question is regarding the determination of the mass of neutrinos which has fundamental implications to both particle physics and cosmology. The neutrino oscillation experiments have established that the neutrinos should have a non-zero mass \citep[e.g.][]{Forero+14,Gonzalez-Garcia+16,Capozzi+17,Salas+17}. However these oscillation experiments are only sensitive to the mass splittings between the neutrino mass eigenstates, and to measure the absolute scale of the neutrino mass other experiments are required. Recently the Karlsruhe Tritium Neutrino (KATRIN) experiment has reported the first direct detection of sub-eV neutrino mass, with an upper limit on the `effective neutrino mass' of 0.8 eV \citep{KATRIN21}. This is based on the kinematic measurements  through  the  observation  of  the  energy  spectrum  of  tritium $\beta$-decay, and is model independent. 

However stronger constraints on the summed neutrino mass ($M_\nu$) can obtained through combining various cosmological probes, as the massive neutrinos leave an imprint on various cosmological observables \citep[e.g.][]{Wong11, LesgourguesPastor12}. But these constraints  have an additional model dependence. Assuming the standard `lambda cold dark matter' ($\Lambda$CDM) model, one of the strongest constraint on the summed neutrino mass has been obtained by combining cosmic microwave background (CMB, \citetalias{Planck2018}), baryonic acoustic oscillation, and redshift-space galaxy clustering \citepalias{eBOSS20} to obtain an upper limit of $M_\nu < 0.102$ eV. However by considering extensions of the standard cosmological  model, the upper limit becomes  less  stringent \cite[e.g.][]{Vagnozzi+18,Choudhury+20}. The community can expect the summed neutrino mass to be measured with increased precision from cosmological probes in the foreseeable future with the advent of next generation of CMB surveys [e.g. the Simons Observatory\footnote{\url{https://simonsobservatory.org/}} \cite[SO,][]{SO}, CMB-S4\footnote{\url{https://cmb-s4.org/}} \citep{CMBS4}], and the stage IV galaxy redshift surveys [e.g. the `Dark Energy Spectroscopic Instrument'\footnote{\url{https://www.desi.lbl.gov/}} \cite[DESI,][]{DESI16}, Euclid\footnote{\url{https://www.euclid-ec.org/}} \citep{Euclid11} and Nancy Grace Roman space telescope\footnote{\url{https://roman.gsfc.nasa.gov/}} \citep{WFIRST15}]. 

Currently the $M_\nu$ constraints from galaxy clustering are mainly obtained using two-point statistics, i.e. the power spectrum in Fourier space or the two-point correlation function in the configuration space. The impact of massive neutrinos on the two-point clustering statistics has been studied quite extensively using $N$-body simulations both in real- \citep[e.g.][]{Saito+08,Wong08,Castorina+15} and redshift-space \citep[e.g.][]{Navarro+18,Garcia+19}. However they are affected by the $M_\nu$-$\sigma_8$ degeneracy, and thus acts as a limitation in measuring the summed neutrino mass.  The three-point clustering statistics in Fourier space (i.e. the bispectrum) has been shown to break this degeneracy \citep{Hahn+20,HahnVilla20}. In addition it has been shown that the three-point cluster statistics contains additional cosmological information compared to its two-point counterpart, and thus able to obtain substantial improvements on  constraining other cosmological parameters also \citep[e.g.][]{YankelevichPorciani18, ChudaykinIvanov19, Gualdi+20, Agarwal+20, Samushia+21}. Currently there are efforts to understand the possibility of constraining summed neutrino mass using various summary statistics, among others, like the one-point probability distribution function of the matter density \citep[e.g.][]{Uhlemann+20} and the void size function \citep[e.g.][]{Bayer+21}.

Another avenue is to use velocity statistics like the mean pairwise velocity which provides a complementary view to the clustering information, either through the peculiar velocity surveys or the kinetic Sunyaev--Zeldovich (kSZ) effect. \cite{Mueller+15b} has shown that the mean pairwise velocity can utilised to constrain the summed neutrino mass, and \cite{Kuruvilla+20} has studied its interplay between the baryonic feedback and the summed neutrino mass effects at nonlinear separation scales. Furthermore the three-point mean relative velocity statistics is able to obtain stronger constraints on the summed neutrino mass when compared to the mean pairwise velocity \citep{KuruvillaAghanim21}. However the growth rate measurement from the kSZ effect of the CMB are degenerate with the optical depth \cite[e.g.][]{KeislerSchmidt13,Battaglia16,Flender+17} and this degeneracy acts as a limitation in measuring cosmological parameters \cite[e.g.][]{Smith+18}, which is commonly referred to as the optical depth degeneracy. It has been suggested that the use of fast radio bursts can be used to break this degeneracy \citep{Madhavacheril+19}. In this paper we will develop a new statistic which is independent of the optical depth using the first moment of the three-point relative velocities, i.e. the mean relative velocities between pairs in a triplet, and thus circumvent the problem of optical depth as a limitation factor in kSZ experiments. This forms one of the main goals of this paper.

The remainder of this work is structured as follows. In Sect.~\ref{sec:cosmoksz}, we describe the newly introduced summary statistic based on three-point mean relative velocities. The Quijote suite of simulation, which we use in this work is introduced briefly in Sect.~\ref{sec:sims}. The information content in the velocity statistics is studied using the Fisher-matrix formalism, which is briefly summarised in Sect.~\ref{sec:fisher}.  Our results are presented in Sect.~\ref{sec:results}, and we finally conclude in Sect.~\ref{sec:conclusions}.

\section{Cosmology using the kSZ effect}
\label{sec:cosmoksz}
\subsection{Kinetic Sunyaev--Zeldovich effect}
\label{sec:kSZ}

As the CMB photons interact with the free electrons of hot ionised gas along the line-of-sight (LOS), the apparent CMB temperature changes. This is due to the fact that there is a transfer of energy from electrons to the resulting scattered photons as the electrons have a significantly higher kinetic energy than the photons. In this work, we focus on the secondary effect which is known as  the kinetic Sunyaev--Zeldovich \citep[kSZ;][]{SZ72,SZ80} which arises if the scattering medium is moving relative to the Hubble flow. The fractional temperature fluctuation caused due to kSZ is
\begin{align}
	\left.\frac{\Delta T(\hat{n})}{T_{\mathrm{cmb}}}\right|_{\mathrm{kSZ}} &= -  \int \mathrm{d}l\ \sigma_{\mathrm{T}} \left(\frac{\bm{v}_{\mathrm{e}}\cdot\hat{n}}{c}\right) n_{\mathrm{e}} \, , \nonumber \\
	& = -\tau \left(\frac{\bm{v}_{\mathrm{e}}\cdot\hat{n}}{c}\right) \, ,
	\label{eq:ksz}
\end{align}
where $\sigma_{\mathrm{T}}$ is the Thomson scattering cross-section, $T_{\mathrm{cmb}}$ is the CMB temperature, $c$ is the speed of light, $\bm{v}_{\mathrm{e}}$ is the peculiar velocity of free electrons, and $n_{\mathrm{e}}$ is the physical free electron number density. The integral $\int \mathrm{d}l$ is computed along the LOS which is given by $\hat{n}$. The optical depth is defined as $\tau = \int \mathrm{d}l\ \sigma_{\mathrm{T}}\ n_{\mathrm{e}}$, i.e. the integrated electron density.  

The kSZ signal detection is challenging because of its small amplitude and its spectrum being identical to that of primary CMB temperature fluctuations. One of the approaches to detect the kSZ signal is to employ the pairwise statistic  \cite[e.g.][]{Hand+12,Carlos+15,Planck16,Schaan+16,Soergel+16,Bernardis+17,Li+17,Calafut+21,Chen+21}. There have been evidences of kSZ signal using other techniques also \cite[e.g.][]{Carlos+15,Schaan+16,Nguygen+20,Tanimura+21,Chaves-Montero+21, Schaan+21}.

In the case of kSZ pairwise signal the temperature acts as a proxy for the peculiar velocity, and as such it probes the optical depth weighted pairwise velocity \cite[e.g.][]{Hand+12,Soergel+18}
\begin{equation}
    \frac{\langle \Delta T^{\mathrm{kSZ}}(r_{12})\rangle}{T_{\mathrm{cmb}}} \simeq -\tau \frac{\bar{w}(r_{12})}{c}\,,
    \label{eq:ksz-pair}
\end{equation}
where $\langle \Delta T^{\mathrm{kSZ}}_{12} \rangle$ is the mean temperature difference between the objects `1' and `2', and $\bar{w}(r_{12})$ is the mean radial component of the pairwise velocity which can be defined in the single streaming regime as follows
\begin{align}
\langle \bm{w}_{12}|\bm{r}_{12} \rangle_\mathrm{p}   = &\ \displaystyle \frac{\langle(1+\delta_{1})(1+\delta_{2}) (\bm{v}_{2}-\bm{v}_{1})\rangle}{\langle (1+\delta_{1})(1+\delta_{2})\rangle} \, ,
 \label{eq:mean-radial-velocity-two}
 \end{align}
where $\delta_i\equiv\delta(\bm{x}_i)$ is the density contrast, $\bm{v}_i\equiv \bm{v}(\bm{x}_i)\equiv \bm{u}(\bm{x}_i)/aH$ is the normalised peculiar velocity, $a$ is the scale factor, and $H$ is the Hubble constant.  Using perturbation theory at leading order (LO), the mean radial matter pairwise velocity can be written as \citep[e.g.][]{Fisher95, Juszkiewicz+98, ReidWhite11}
\begin{equation}
\langle \bm{w}_{12}|\bm{r}_{12} \rangle_{\mathrm p}  = \bar{w}(r_{12})\,\hat{\bm{r}}_{12}
 \simeq \displaystyle - \frac{f}{\pi^2} \, \hat{\bm{r}}_{12} \int_0^{\infty} k \,j_1(k r_{12})\,P(k)\, \rd k \, ,
 \label{eq:mean-radial-velocity-two-theory}
\end{equation}
where $\hat{\bm{r}}_{12}$ is the unit vector along the pair `12', the subscript p implies that the averages are computed over all pairs with separation $r_{12}$, $P(k)$ denotes the linear matter power spectrum, and $j_1(x) = \sin (x)/x^2- \cos (x)/x$. It should be noted that Eq.~(\ref{eq:ksz-pair}) assumes that there is no correlation between optical depth and velocity field. Following Eqs.~(\ref{eq:ksz-pair}) and (\ref{eq:mean-radial-velocity-two-theory}), we can see that
\begin{equation}
    \Delta T^{\mathrm{kSZ}} \propto \tau f \sigma^2_8 \, 
\end{equation}
and thus implying that the growth rate measurement from the pairwise kSZ is perfectly degenerate with optical depth \citep{KeislerSchmidt13}. Here we have presented the argument for the matter, but in observations there will be an additional bias dependence entering in the above equation.

\subsection{New statistics based on mean relative velocity between pairs in a triplet}
\label{sec:velsta}

In the previous section, we mentioned about the mean relative velocity between two tracers (i.e between a pair) or the mean pairwise velocity. However this can be generalised to the case of three tracers, in which we can consider two mean relative velocity between pairs in a triplet with separations $\triangle_{123}=(r_{12},r_{23},r_{31})$:
(i) $\langle \bm{w}_{12}|\triangle_{123} \rangle_\mathrm{t}$, and (ii)
$\langle \bm{w}_{23}|\triangle_{123} \rangle_\mathrm{t}$.
The subscript t here implies that the averages are computed over all triplets with separations $(r_{12},r_{23},r_{31})$. Similar to Eq.~(\ref{eq:mean-radial-velocity-two}) in the single stream fluid approximation, the mean relative velocity between pair 12 in a triplet can be written as \citep{KuruvillaPorciani20}
\begin{align}
\langle \bm{w}_{12}|\triangle_{123} \rangle_\mathrm{t}   = &\ \displaystyle \frac{\langle(1+\delta_{1})(1+\delta_{2}) (1+\delta_{3})(\bm{v}_{2}-\bm{v}_{1})\rangle}{\langle (1+\delta_{1})(1+\delta_{2}) (1+\delta_{3})\rangle} \nonumber \\
 \simeq &\ \langle \delta_{1}\bm{v}_{2} \rangle - \langle \delta_{2} \bm{v}_{1} \rangle + \langle \delta_{3}\bm{v}_{2} \rangle - \langle \delta_{3} \bm{v}_{1} \rangle\nonumber \\
 =&\ \bar{w}(r_{12})\,\hat{\bm{r}}_{12}-\frac{1}{2}\left[
 \bar{w}(r_{23})\,\hat{\bm{r}}_{23}+\bar{w}(r_{31})\,\hat{\bm{r}}_{31}\right]
 \, .
 \label{eq:mean-radial-velocity-three}
 \end{align}

The three-point mean relative velocity statistics can be composed into both its radial ($R_{ij}$) and transverse ($T_{ij}$) component in the plane of the triangle defined by the particles. This is in contrast to the mean pairwise velocity for which the transverse component is zero. In the case of $ \langle \bm{w}_{12}|\triangle_{123} \rangle_\mathrm{t}$, it is as follows
\begin{align}
   \langle \bm{w}_{12}|\triangle_{123} \rangle_\mathrm{t}  & = \langle \bm{w}_{12}\cdot \hat{\bm{r}}_{12} |\triangle_{123} \rangle_{\mathrm t}\,\hat{\bm{r}}_{12} +
   \langle \bm{w}_{12}\cdot \hat{\bm{t}}|\triangle_{123} \rangle_{\mathrm t}\, \hat{\bm{t}}\nonumber\\
&   =R_{12}(\triangle_{123})\,\hat{\bm{r}}_{12}+T_{12}(\triangle_{123})\,\hat{\bm{t}}\;,
   \label{eq:decomposition}
\end{align}
where $\hat{\bm{t}}=(\hat{\bm{r}}_{23}-\cos\chi\,\hat{\bm{r}}_{12})/\sin\chi$, $\hat{\bm{r}}_{23}$ is the unit vector along the pair `23', and $\chi= \arccos(\hat{\bm r}_{12} \cdot \hat{\bm r}_{23})$. In this work, we make use of only the radial component, and for the pair 12 in the triplet it  can be written as 
 \begin{align}
 R_{12}(\triangle_{123})=
 \bar{w}(r_{12})&-\frac{1}{2}\Bigg[
 \bar{w}(r_{23})\,\cos \chi \nonumber \\
 & -\bar{w}(r_{31})\,\frac{r_{12}+r_{23}\cos\chi}{\sqrt{r_{12}^2+r_{23}^2+2r_{12}r_{23}\cos\chi}}\Bigg]\;.
 \label{eq:R12_triangle}
 \end{align}

 \noindent Similarly, the mean radial relative velocity between the pair 23 in $\triangle_{123}$ can be written as
\begin{align}
 R_{23}(\triangle_{123})=
 \bar{w}(r_{23})&-\frac{1}{2}\Bigg[
 \bar{w}(r_{12})\,\cos \chi \nonumber \\
 & -\bar{w}(r_{31})\,\frac{r_{23}+r_{12}\cos\chi}{\sqrt{r_{12}^2+r_{23}^2+2r_{12}r_{23}\cos\chi}}\Bigg]\;.
 \label{eq:R23_triangle}
 \end{align}

\noindent Similar to Eq.~(\ref{eq:ksz-pair}), the three-point mean relative temperature difference from kSZ can be written down as
\begin{equation}
    \frac{\Delta T^{\mathrm{kSZ}}_{ij}(\triangle_{123})}{T_{\mathrm{cmb}}} \simeq -\tau \frac{R^{\mathrm{h}}_{ij}(\triangle_{123})}{c}\,,
    \label{eq:ksz-triplet}
\end{equation}
where $R^{\mathrm{h}}_{ij}(\triangle_{123})$ respresents the three-point mean relative velocity statistics for haloes (biased tracers), and to first approximation it can be written down as linear bias term times $R_{ij}(\triangle_{123})$ \citep{KuruvillaAghanim21}.  Based on the radial mean relative velocities between pairs in a triplet, we can introduce a new ratio statistic 
\begin{align}
    \mathcal{R}(\triangle_{123}) &= \frac{\langle \bm{w}_{12}\cdot \hat{\bm{r}}_{12} |\triangle_{123} \rangle_{\mathrm t}}{\langle \bm{w}_{23}\cdot \hat{\bm{r}}_{23} |\triangle_{123} \rangle_{\mathrm t}}  \, ,  \label{eq:ratio-new-statistics}
\end{align}
which tells us how quickly the average infall velocity of pair 12 is in comparison to the average infall velocity of pair 23 for a specific triangular configuration $\triangle_{123}$. On linear scales, using perturbation theory at LO, $\mathcal{R}(\triangle_{123})$ can be written as the ratio between Eqs.~(\ref{eq:R12_triangle}) and (\ref{eq:R23_triangle}) 
\begin{align}
    \mathcal{R}(\triangle_{123})&=\frac{R_{12}(\triangle_{123})}{R_{23}(\triangle_{123})} = \frac{R^{\mathrm{h}}_{12}(\triangle_{123})}{R^{\mathrm{h}}_{23}(\triangle_{123})} \equiv \frac{\Delta T^{\mathrm{kSZ}}_{12}(\triangle_{123})}{\Delta T^{\mathrm{kSZ}}_{23}(\triangle_{123})} \,  .
    \label{eq:ratio-new-statistics-lo}
\end{align}
The above introduced statistic is thus independent of optical depth, $\sigma_8$ and linear bias. In the following sections, we take a detailed look at whether the Ansatz of $\sigma_8$ and linear bias independence holds up. Additionally, we study the cosmological information content in  $\mathcal{R}(\triangle_{123})$.

\section{Data and analysis}
\subsection{Quijote simulation suite}
\label{sec:sims}

In this work, we make use of the Quijote\footnote{\url{https://quijote-simulations.readthedocs.io/}} \citep{Quijote20} suite of simulations, which was run using the tree-PM code \textsc{gadget}-3 \citep{Springel05}. Spanning more than a few thousand cosmological models, it contains 44,100 \textit{N}-body simulations. These simulations have a box length of $1\,h^{-1}\,\mathrm{Gpc}$, and tracks the evolution of $512^3$ cold dark matter (CDM) particles. The initial conditions (ICs) were generated at redshift $z=127$ using the second-order Lagrangian perturbation theory. The fiducial cosmological parameters (assuming zero summed neutrino mass) for the simulation is as follows: the total matter density: $\Omega_{\mathrm{m}}=0.3175$, the baryonic matter density: $\Omega_{\mathrm{b}}=0.049$, the primordial spectral index of the density perturbations: $n_{\mathrm{s}}=0.9624$, the amplitude of the linear power spectrum on the scale of $8\ h^{-1}\mathrm{Mpc}$: $\sigma_8=0.834$,  and the present-day value of the Hubble constant: $H_0\equiv H(z=0)=100\, h\,\mathrm{km}\,\mathrm{s}^{-1}\mathrm{Mpc}^{-1}$ with $h=0.6711$. This is broadly consistent with the Planck 2018 result (\citetalias{Planck2018}). The suite consists of 15,000 random realisations for the fiducial cosmology. For the purpose of calculating derivatives, Quijote provides a set of 500 random realisations wherein only one parameter is varied with respect to the the fiducial cosmology. The variations are as follows: \{$\Omega^+_{\mathrm{m}}, \Omega^-_{\mathrm{m}}, \Omega^+_{\mathrm{b}}, \Omega^-_{\mathrm{b}}, n^+_{\mathrm{s}}, n^-_{\mathrm{s}}, \sigma^+_{8}, \sigma^-_{8}\} = \{0.3275, 0.3075, 0.051, 0.047, 0.9824, 0.9424, 0.849, 0.819\}$ and $\{h^+, h^-\} = \{0.6911,$ $0.6511\}$.

In addition the suite also provides 500 realisations for three massive neutrino cosmology, where the summed neutrino masses are 0.1, 0.2, and 0.4 eV respectively. The initial conditions for these simulations were produced using the Zeldovich approximation (ZA), and has $512^3$ neutrino particles in addition to the CDM particles. To compute the numerical derivatives with respect to massive neutrinos, the Quijote suite provides an addition 500 random realisations for the fiducial cosmology, in which the ICs were also generated using ZA.

In this work, we use halo catalog data from 22,000 \textit{N}-body simulations of the Quijote suite. These halos  were identified using a friends-of-friends algorithm. We selected halos that have a halo mass $M_\mathrm{h} > 5 \times 10^{13}\ h^{-1}\mathrm{M}_\odot$ (corresponding to groups and clusters of galaxies) at $z=0$ which gives a mean number density of $\bar{n} \sim 0.92 \times 10^{-4}\,h^3\,\mathrm{Mpc}^{-3}$ for the reference simulations. Additionally in the case of (i) fiducial cosmology, and (ii) for variations in $\sigma_8$ (both $\sigma^+_{8}$ and $\sigma^-_{8}$), we use 30 realisations of the particle data (randomly down-sampled to $100^3$ particles) to compute $\mathcal{R}(\triangle_{123})$. 

\subsection{Fisher-matrix formalism}
\label{sec:fisher}

To quantify the error estimates on the cosmological parameters, we use the Fisher-matrix formalism which can be defined as \citep[e.g.][]{Tegmark+1997, Heavens09, Verde10}
\begin{equation}
    F_{\alpha \beta} = \left\langle -\frac{\partial^2\ln{\mathcal{L}}}{\partial \theta_\alpha \partial \theta_\beta} \right\rangle \, ,
    \label{eq:fisher_definition}
\end{equation}
where $\theta_\alpha$ and $\theta_\beta$ are two of the cosmological model parameters, and $\mathcal{L}$ is the likelihood of the data given a model. Assuming a Gaussian likelihood, we can write the Fisher information matrix as
\begin{equation}
    F_{\alpha \beta} = \frac{\partial \mkern 1mu \boldsymbol{\mathcal{R}}}{\partial \theta_\alpha} \cdot \hat{\mathbf{C}}^{-1} \cdot \frac{\partial \mkern 1mu \boldsymbol{\mathcal{R}}^\mathsf{T}}{\partial \theta_\beta} \, ,
    \label{eq:fisher_reduced}
\end{equation}
where $\boldsymbol{\mathcal{R}}$ represents the data vector for the ratio statistic we introduced in Eq.~(\ref{eq:ratio-new-statistics}), and $\hat{\mathbf{C}}^{-1}$ is the precision matrix (i.e. the inverse covariance matrix). It should be noted that in the definition of $F_{\alpha\beta}$, we have neglected a term which appears due to the cosmology dependence of the covariance matrix. However the correction has been shown to have a negligible effect \citep{Kodwani+19}. We compute the covariance matrix of $\mathcal{R}$ directly from the simulations as follows
\begin{equation}
    \widetilde{\mathbf{C}} = \frac{1}{N_{\mathrm{sims}}-1}\sum_{i=1}^{N_{\mathrm{sims}}} \left(\boldsymbol{\mathcal{R}}_i-\overline{\boldsymbol{\mathcal{R}}}\right)\left(\boldsymbol{\mathcal{R}}_i-\overline{\boldsymbol{\mathcal{R}}}\right)^\mathsf{T} \, ,
    \label{eq:covariancematrix}
\end{equation}
where $\overline{\boldsymbol{\mathcal{R}}} = N_{\mathrm{sims}}^{-1}\sum_{i=1}^{N_{\mathrm{sims}}} \boldsymbol{\mathcal{R}}_i$, and $N_{\mathrm{sims}}$ denotes the total number of simulations used to compute the covariance matrix (in this work $N_{\mathrm{sims}}=15,000$). While Eq.~(\ref{eq:covariancematrix}) gives an unbiased estimate of the covariance matrix, its inversion leads to a biased estimate of the precision matrix. This however can be statistically corrected by applying a multiplicative correction factor to the precision matrix \citep{Kaufmann67,Anderson03,Hartlap+07}
\begin{equation}
    \hat{\mathbf{C}}^{-1} = \frac{N_{\mathrm{sims}}-N_{\mathrm{bins}}-2}{N_{\mathrm{sims}}-1}\ \widetilde{\mathbf{C}}^{-1} \, ,
\end{equation}
where $N_{\mathrm{bins}}$ is the number of bins in $\mathcal{R}$.

We numerically compute the derivatives required to construct the Fisher information matrix using the Quijote suite of simulations, which provides 500 realisations where only one cosmological parameter is varied while the rest are fixed at its fiducial value. Thus in the case when the model parameters are one of the follows: $\theta \equiv \{\Omega_{\mathrm{m}}, \Omega_{\mathrm{b}}, h, n_{\mathrm{s}}, \sigma_{8}\}$, we make use of the central difference approximation to compute the derivative numerically
\begin{equation}
    \frac{\partial \mkern 1mu \boldsymbol{\mathcal{R}}}{\partial \theta} \simeq \frac{\boldsymbol{\mathcal{R}}(\theta+\mathrm{d}\theta)-\boldsymbol{\mathcal{R}}(\theta-\mathrm{d}\theta)}{2\ \mathrm{d}\theta} \, .
\end{equation}

In the case of the neutrino mass, the fiducial value is 0.0 eV and it cannot have negative values, hence we obtain the partial derivative using
\begin{equation}
    \frac{\partial \mkern 1mu  \boldsymbol{\mathcal{R}}}{\partial M_\nu} \simeq \frac{-\boldsymbol{\mathcal{R}}(M_\nu=0.4)+4\boldsymbol{\mathcal{R}}(M_\nu=0.2) - \boldsymbol{\mathcal{R}}(M_\nu=0)}{0.4} \, .
\end{equation}
Thus we utilise two sets of massive neutrino simulations from Quijote, with $M_\nu = 0.2$ eV and $M_\nu = 0.4$ eV for the Fisher information matrix. However the initial condition of the simulations with the massive neutrinos were generated using ZA. To be consistent to compute the partial derivative, we make use of another 500 realisations of fiducial cosmology (with $M_\nu = 0$ eV) simulation in which the initial conditions were also generated using ZA.

\section{Results}
\label{sec:results}

\begin{figure}
  \centering
  \includegraphics[scale=0.56]{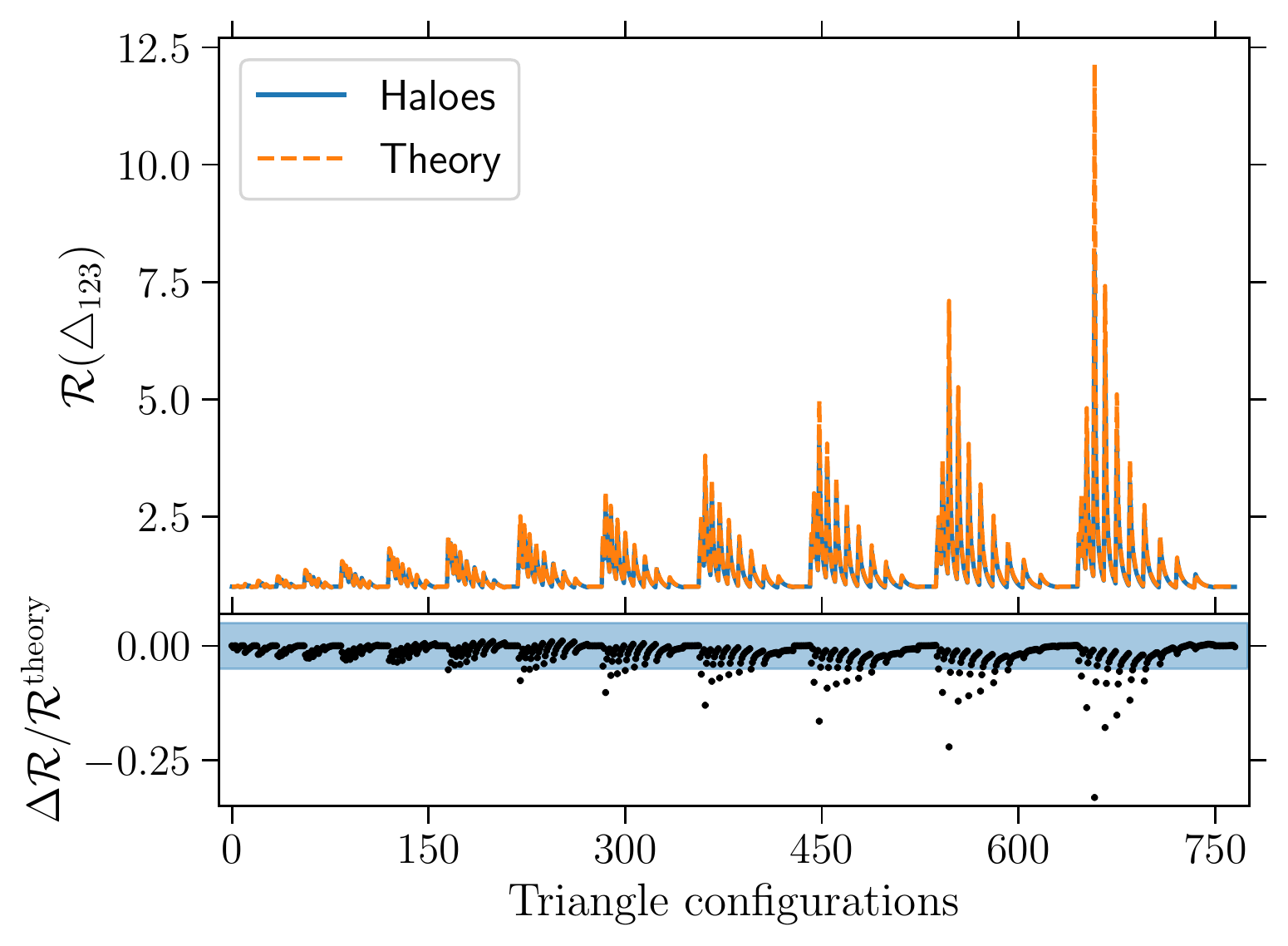}
  \caption{Top: comparing theoretical prediction (orange dashed line) for $\mathcal{R}(\triangle_{123})$ using perturbation theory at LO against the direct measurement (blue solid line) from the halo catalogs of the Quijote suite of simulation. Bottom: residual showing the deviation of the theoretical prediction from the direct measurement from the simulations. The blue shaded region denotes the 5\% region. The triangle configuration go from the smallest being $\{(40,45), (40,45), (40,45)\} \,h^{-1}\mathrm{Mpc}$ to the largest which corresponds to $\{(115,120), (115,120), (115,120)\} \,h^{-1}\mathrm{Mpc}$.}
  \label{fig:theory}
\end{figure}

In Fig.~\ref{fig:theory} we show the direct measurements of $\mathcal{R}(\triangle_{123})$ from the 15,000 reference halo catalogs (solid blue line), and compare it against the LO prediction (dashed orange line). We consider all triangular configurations with  $r_\mathrm{min} \in (40, 45)$ and $r_\mathrm{max} \in (115, 120)$, and such that $r_{12} \geq r_{23} \geq r_{31}$. All the separation scales have a bin width of 5 $h^{-1}\mathrm{Mpc}$. It thus corresponds to a total of 766 triangular configurations, spanning from configuration `0' being the smallest (i.e. $\triangle_{123} \in \{(40,45), (40,45), (40,45)\} \,h^{-1}\mathrm{Mpc}$) to configuration `765' being the largest ($\triangle_{123} \in \{(115,120), (115,120), (115,120)\} \,h^{-1}\mathrm{Mpc}$).  One can see from Eqs.~(\ref{eq:R12_triangle}) and (\ref{eq:R23_triangle}) that the mean three-point relative velocities, $R_{12}$ and $R_{23}$, will be equal to each other when $r_{12} = r_{23}$, irrespective of the length of the third side. This is directly visible in Fig.~\ref{fig:theory}, where $\mathcal{R}=1$ when this condition is met. When comparing the theoretical predictions, we see that it is overall accurate within 4--5\% for configurations with all separation lengths greater than 55 $h^{-1} \mathrm{Mpc}$. As expected when the separation length decreases, the fidelity of the LO prediction also decreases with the maximum deviation at about 27\% for the triangular configuration $\{(100,105),(50,55),(50,55)\}$ $h^{-1}\mathrm{Mpc}$. This thus motivates us to directly measure $\mathcal{R}$ from the simulations to compute the derivatives for the Fisher information matrix.

\begin{figure}
  \centering
  \includegraphics[scale=0.58]{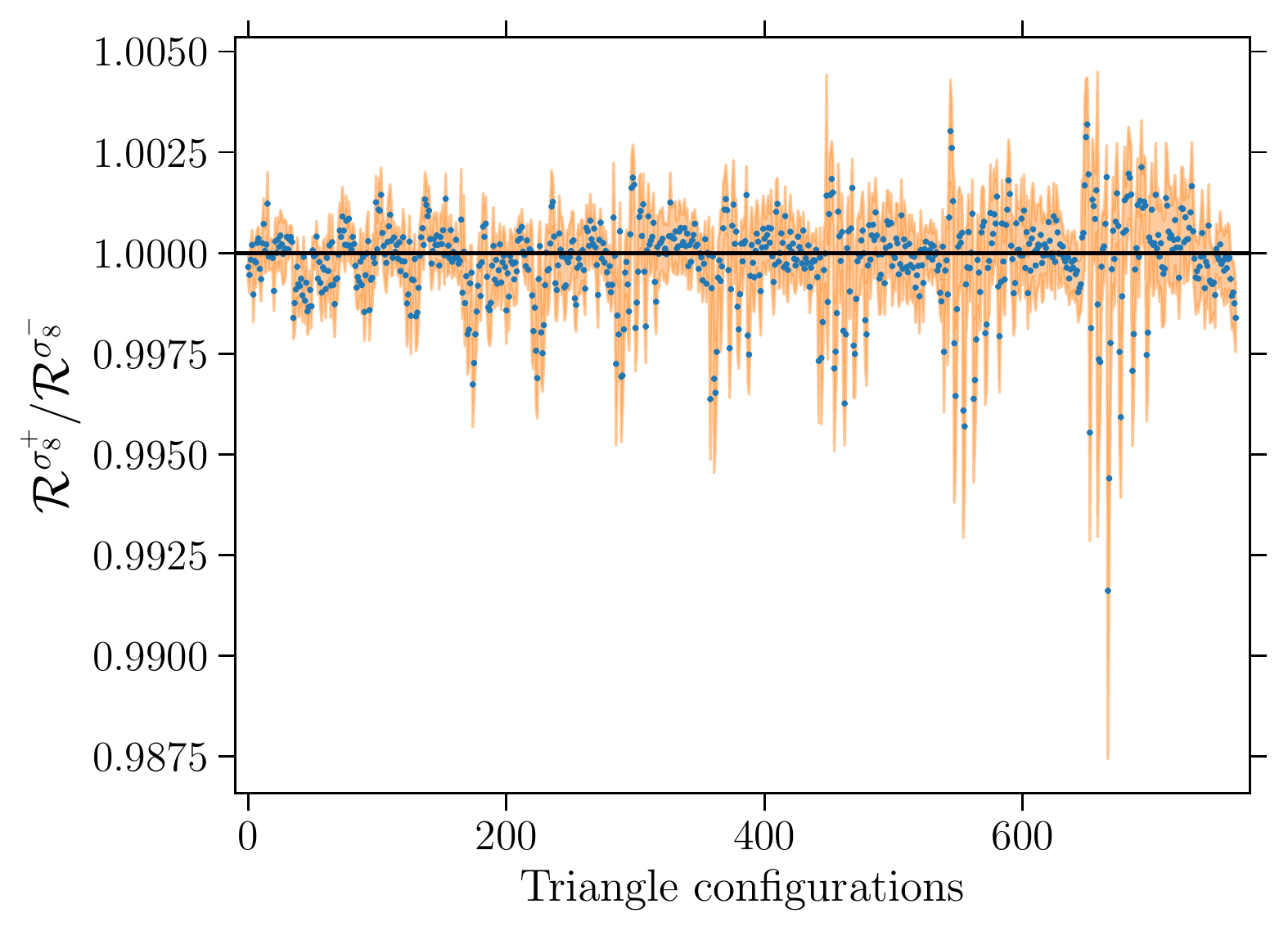}
  \caption{Ratio of $\mathcal{R}$ at $\sigma_8^+=0.849$ to $\mathcal{R}$ at $\sigma_8^-=0.819$. The (blue) solid dot, and the (orange) error bar represents the mean and the relative error on the mean, respectively.} 
  \label{fig:s8_independence}
\end{figure}

\begin{figure}
  \centering
  \includegraphics[scale=0.58]{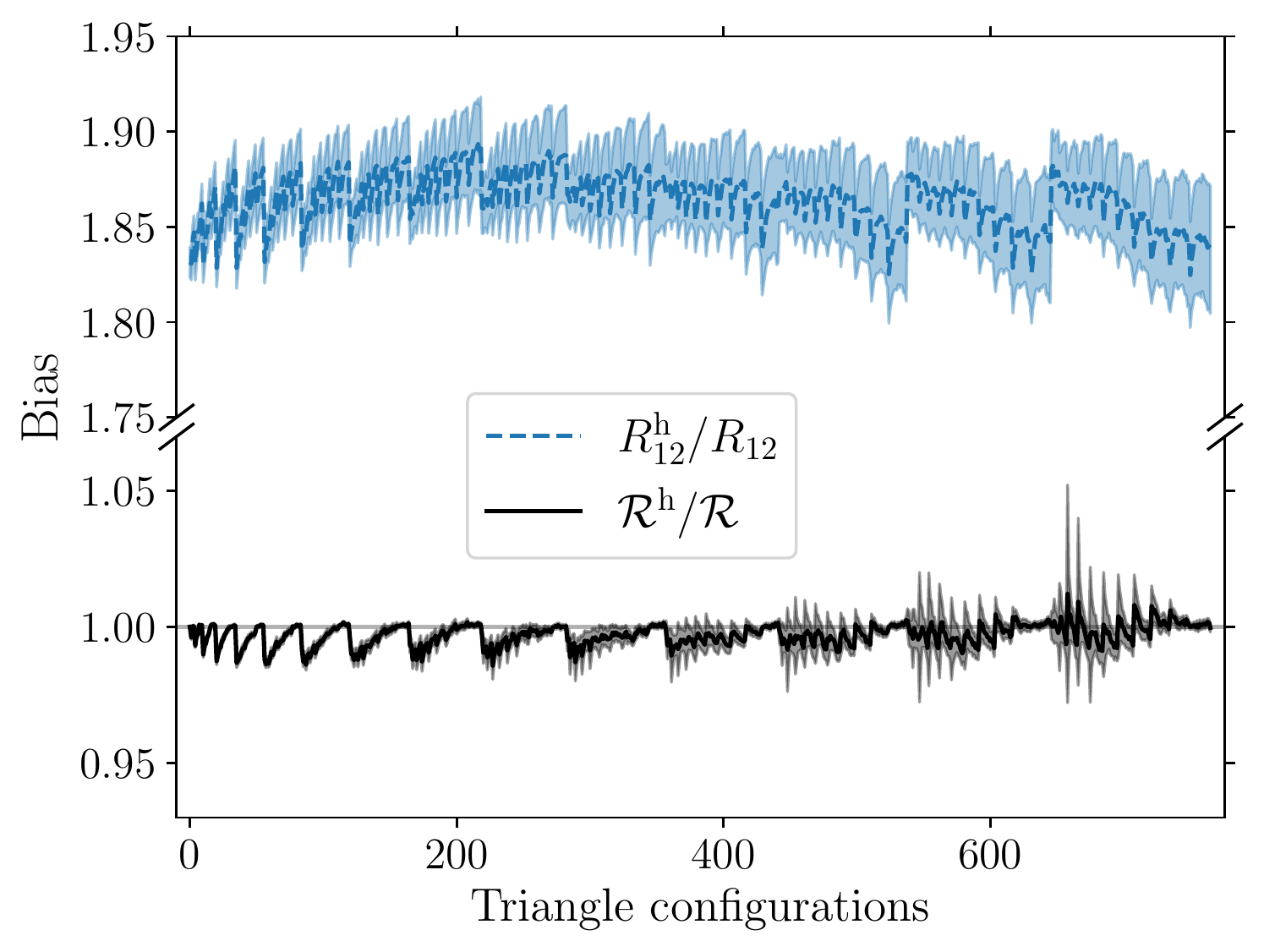}
  \caption{The dashed (blue) line shows the bias for $R_{12}$, i.e. the mean radial relative velocity between pairs 1 and 2 in a triplet. While the solid (black) line shows the (weak) bias dependence of the ratio statistics $\mathcal{R}$, which is equal to one within 1--2$\%$ for all triangular configurations considered here.} 
  \label{fig:halo_bias_independence}
\end{figure}

In addition $\mathcal{R}(\triangle_{123})$ is unaffected by variation in $\sigma_8$ as mentioned earlier in Sect.~\ref{sec:velsta}. To demonstrate this, we compute the ratio of $\mathcal{R}(\triangle_{123})$ for $\sigma^+_8 = 0.849$ and $\sigma^-_8 = 0.819$ using dark matter particles from 30 realisations, and showcase it in Fig.~\ref{fig:s8_independence}. The (blue) dot represents the mean of the measurement, while the scatter is shown using the (orange) error bars. Thus we can conclude that $\mathcal{R}$ is independent of $\sigma_8$. Similarly in Fig.~\ref{fig:halo_bias_independence}, we take a look at the bias dependence of $\mathcal{R}(\triangle_{123})$ (black solid line), and $R_{12}$ (blue dashed line), where we show the ratio of each between the halo and matter component for each of the summary statistics. The bias term for the mean relative velocity between pair `23' in a triplet is similar to $R^{\mathrm{h}}_{12}$, and is thus not shown in the figure. As reported in \cite{KuruvillaAghanim21}, for these triangular configurations (assuming a scale independent bias) it yields a bias factor around 1.85 for $R_{12}$ and $R_{23}$.  For the purpose of computing these ratios in the figure, we used the mean relative velocity information for matter from 30 realisations of the dark-matter only simulations, and for halo we utilised the 15,000 catalogues. The shaded regions represents the $1\sigma$ errors from the propagation of uncertainties of the mean relative velocity statistics for matter, and halo. One can see that on large separation scales the newly introduced statistic (black solid line) is bias independent, while for the smallest triangle configuration there is a very weak dependence of bias when considering the newly introduced statistic. This thus supports the Ansatz presented in Eq.~(\ref{eq:ratio-new-statistics-lo}), where the LO in perturbation theory renders $\mathcal{R}$ to be bias independent on linear scales. For all triangular configurations considered in this work the bias is found to be equal to one within 1--2\%, and hence for the purpose of Fisher matrix formalism we consider $\mathcal{R}$ being independent of a (constant) linear bias term.

\begin{figure}
  \centering
  \includegraphics[scale=0.56]{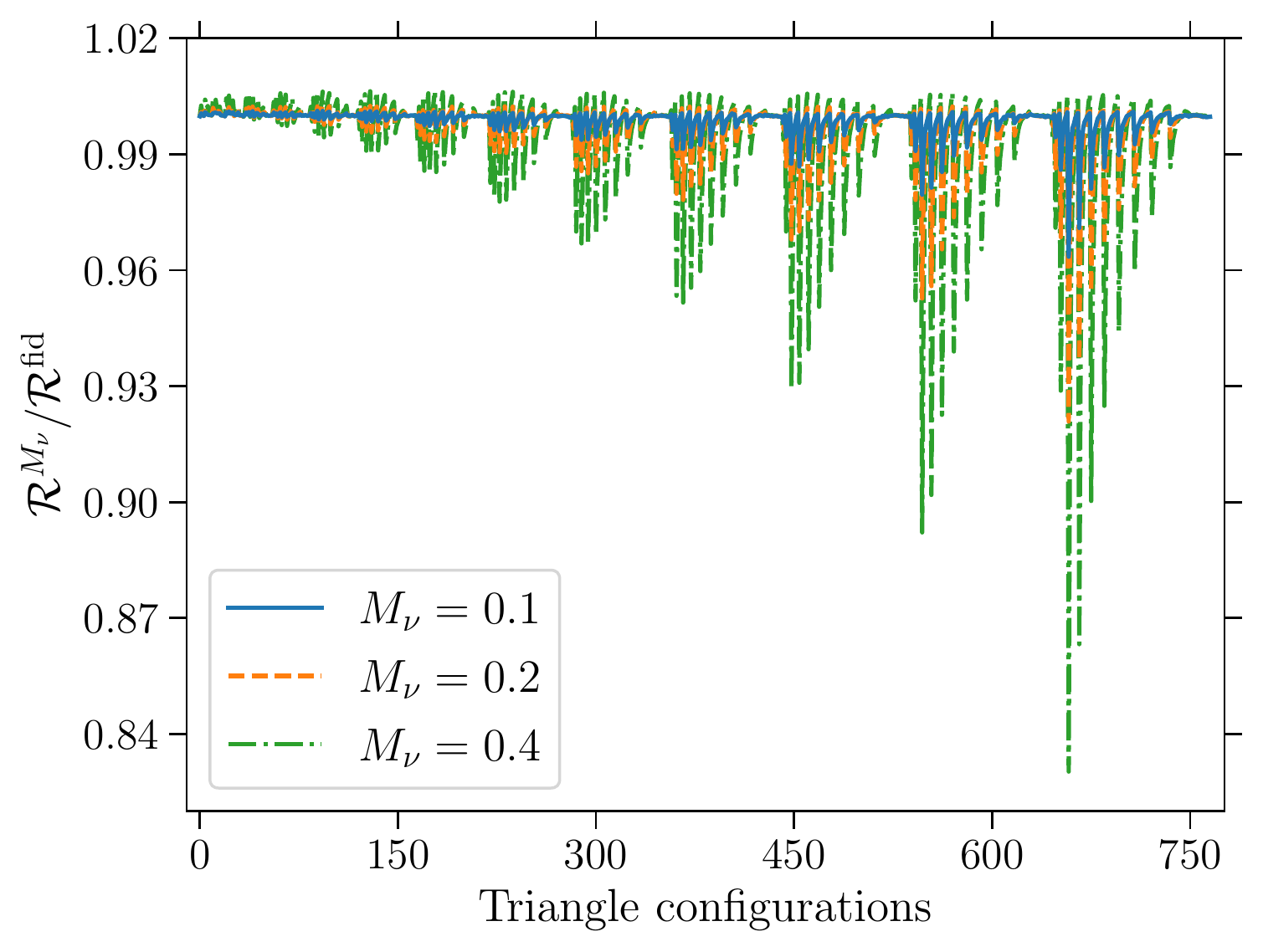}
  \caption{Effect of summed neutrino mass on $\mathcal{R}(\triangle_{123})$, as measured directly from the simulations, when compared to zero neutrino mass fiducial cosmology. The summed neutrino mass considered here are denoted in the legend, and its units are given in eV.} 
  \label{fig:halo_neutrino_mass}
\end{figure}

Since $\mathcal{R}$ is found to be independent of $\sigma_8$ at the scales we are probing (i.e. $r_{\mathrm{min}} \geq 40\ h^{-1}\mathrm{Mpc}$), this renders the statistic at an unique position of being unaffected by the degeneracy in the $M_\nu$-$\sigma_8$ parameter plane. We check the impact of summed neutrino mass on $\mathcal{R}$ utilising three non-zero neutrino mass in Fig.~\ref{fig:halo_neutrino_mass}. The solid (blue) line shows the impact of $M_\nu=0.1$ eV on $\mathcal{R}$ when compared to zero neutrino mass cosmology. Similarly the dashed (orange) and dash-dotted (green) lines showcases the impact of $M_\nu=0.2$ and $M_\nu=0.4$ eV, respectively. As can be seen, when the neutrino mass increases there is a decrease in the infall velocity between a pair in most of the triangular configurations. This is related to the free-streaming of neutrinos as a result of them having large thermal velocities. As a result, below the free-streaming scale neutrinos does not cluster, which further slows down the collapse of the matter in general. This leads to an overall reduction in the growth of overall density perturbations at scales below free-streaming scale, and thus causes a suppression of power on large Fourier modes when looking at the matter power spectrum \citep[e.g.][]{Wong11,LesgourguesPastor12}. When looking at $\mathcal{R}$ for all the configurations we measured, the maximal effect of suppression is seen in the case when $M_\nu=0.4$ eV, and for the triangular configuration $\{(100,105),(50,55),(50,55)\} \ h^{-1}\mathrm{Mpc}$  when compared to the zero neutrino mass cosmology.

\subsection{Cosmological parameters}
\label{sec:cosmoparams}

\begin{figure}
  \centering
  \includegraphics[scale=0.62]{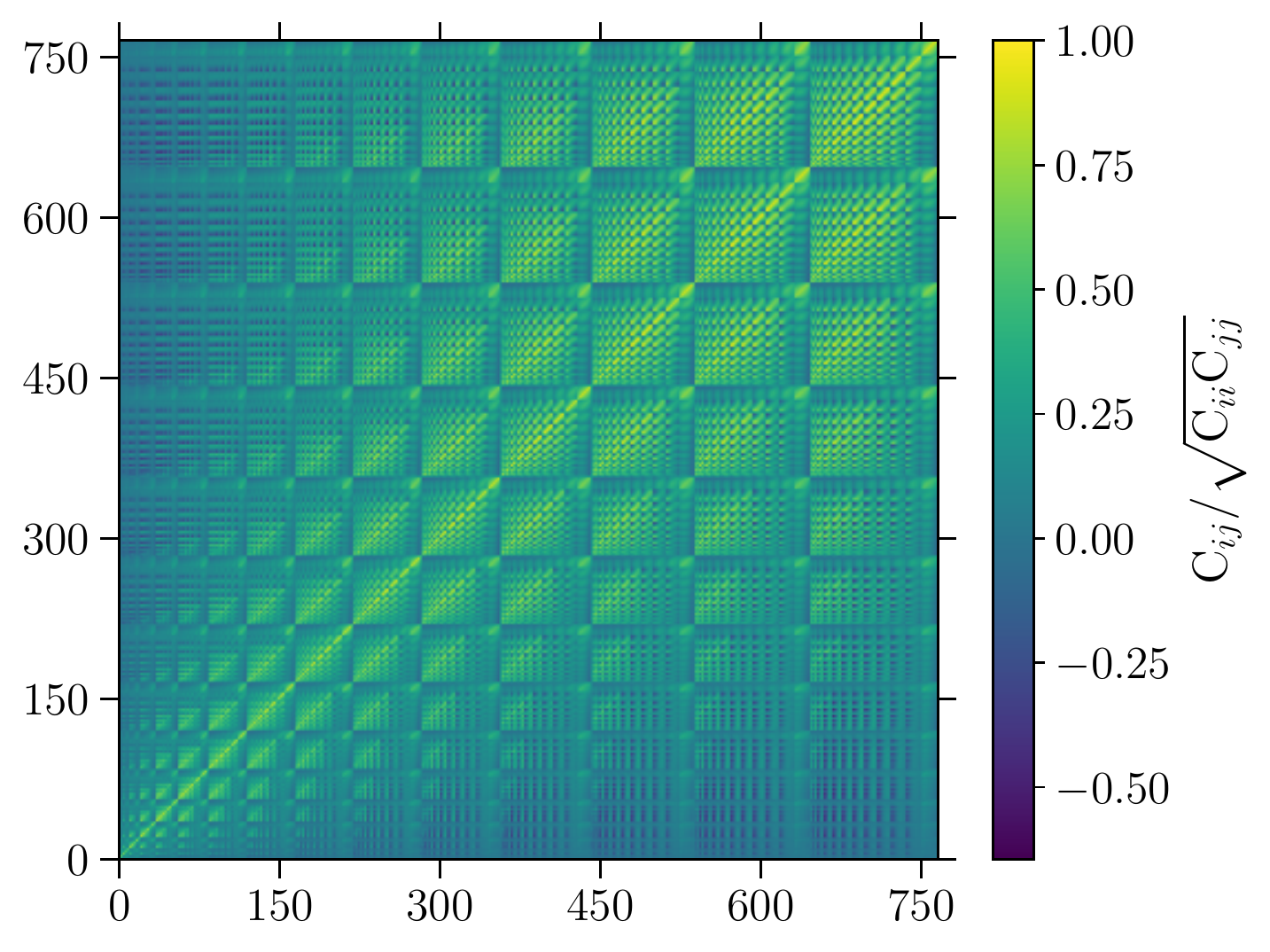}
  \caption{The correlation matrix (i.e. the covariance matrix of $\mathcal{R}$ normalised by its diagonal elements) computed using 15,000 realisations of the Quijote simulations. The triangle configurations are same as in Fig.~\ref{fig:theory}.} 
  \label{fig:correlation_matrix}
\end{figure}
\begin{figure*}
  \centering
  \includegraphics[scale=0.53]{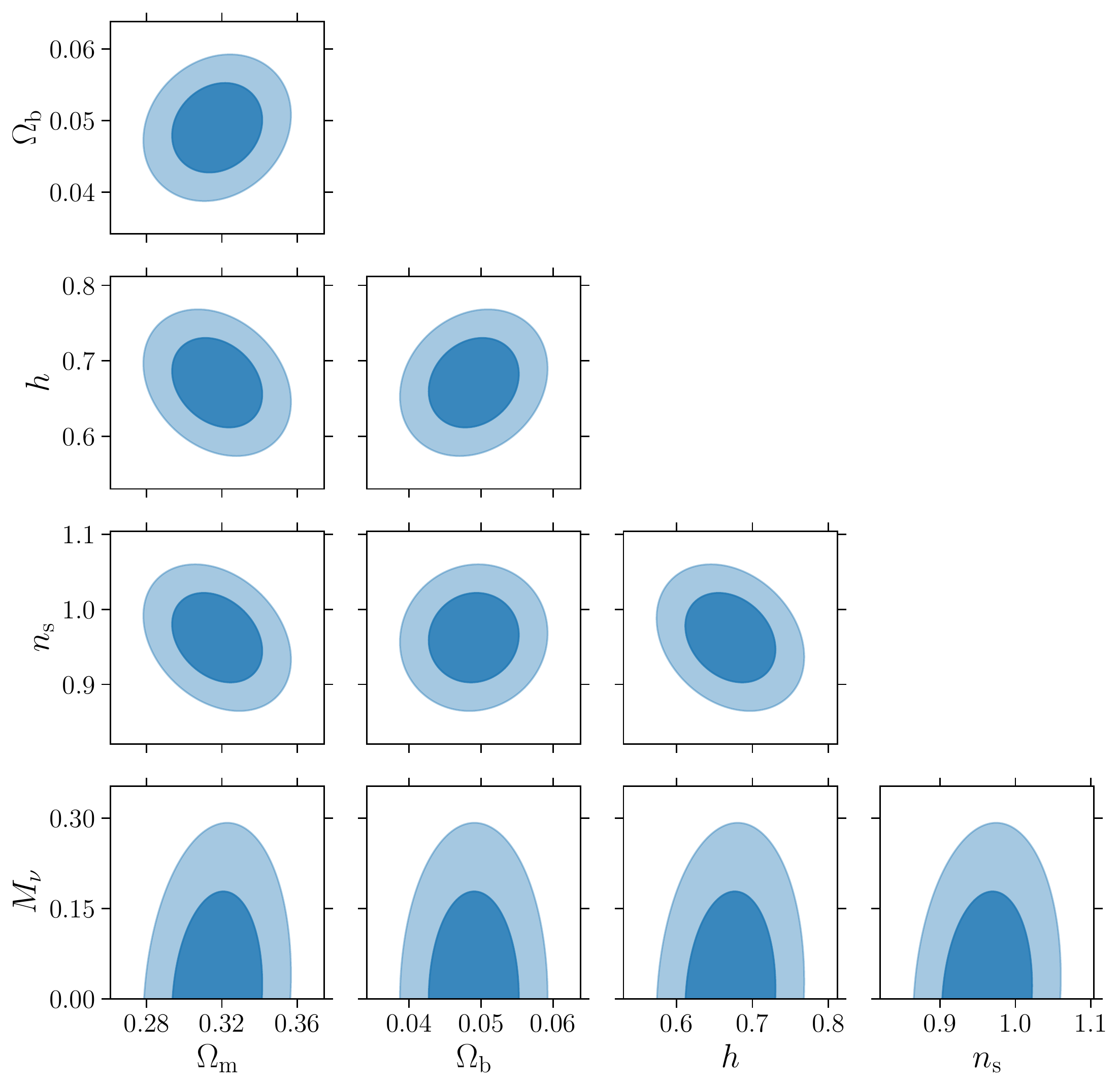}
  \caption{Joint 68.3\% (dark shaded contour) and 95.4\% (light shaded contour) credible region for all the pairs of cosmological model parameters at $z=0$.} 
  \label{fig:fisher_results}
\end{figure*}

We now turn our attention to the information content in the new ratio statistic $\mathcal{R}$, and see its viability in constraining the cosmological model. As discussed in Sect.~\ref{sec:fisher} we achieve this using the Fisher information matrix, and the ingredients for it are the partial derivatives of $\mathcal{R}$ with respect to the cosmological model parameters and the covariance matrix ($\mathbf{C}$). We show case the correlation matrix in Fig.~\ref{fig:correlation_matrix}, which is given as $\mathrm{C}_{ij}/\sqrt{\mathrm{C}_{ii}\mathrm{C}_{jj}}$, wherein the covariance matrix is directly measured from the simulations using 15,000 realisations. We notice the presence of non-diagonal terms in it, being positively correlated at similar triangular configurations while tending to be negatively correlated as the configurations differ substantially. 

We now take a look at the information content in $\mathcal{R}$ using the Fisher information matrix formalism, as defined in Sect.~\ref{sec:fisher}. As mentioned, we have computed both the elements of it directly from the simulations. Since the bias dependence of $\mathcal{R}$ was shown to being very weak even at small scales ($\sim$ 40--50 $h^{-1}\mathrm{Mpc}$) and being independent at large scales ($\geq 80$ $h^{-1}\mathrm{Mpc}$), we do not consider the bias parameter in the Fisher-matrix formalism. Thus the model parameters are $\Omega_{\mathrm{m}}$, $\Omega_{\mathrm{b}}$, $h$, $n_{\mathrm{s}}$, and $M_\nu$. We show the results of our Fisher forecast in Fig.~\ref{fig:fisher_results}, where the dark and light shaded contours denote the 68.3\% and 95.4\% joint credible region for all possible model parameters, respectively. The $1\sigma$ marginalised error for any model parameter $\theta_{\alpha}$ is given by $\sqrt{F_{\alpha\alpha}^{-1}}$, and are as follows for the parameters we considered: $\{\Omega_{\mathrm{m}}, \Omega_{\mathrm{b}}, h, n_{\mathrm{s}},M_\nu\} \equiv \{0.0158, 0.0041, 0.0391, 0.0394, 0.1175\}$.

We compare these constraints with those obtained from the mean pairwise velocity, and the mean relative velocity between pairs in a triplet as reported in \cite{KuruvillaAghanim21}. As a fair comparison, we use the constraints obtained from them using $r_{\mathrm{min}} = 40\ h^{-1}\mathrm{Mpc}$. It presents a factor of improvement of \{6.2, 7.6, 9.8, 12.9, 8.87\} for \{$\Omega_{\mathrm{m}}$, $\Omega_{\mathrm{b}}$, $h$, $n_{\mathrm{s}}$, $M_\nu$\}, respectively over the mean pairwise velocity. However when comparing against the constraints from $R^{\mathrm{h}}_{12}(\triangle_{123})+R^{\mathrm{h}}_{23}(\triangle_{123})$ (i.e. the mean three-point relative velocities), $\mathcal{R}$ has its constraining power shrunk by a factor of 1.34--1.44 for all the model parameters. This could also due to the fact that the length of the data vector is twice in $R^{\mathrm{h}}_{12}+R^{\mathrm{h}}_{23}$ when compared to $\mathcal{R}$. This shrinkage in constraining power was also seen when considering $R^{\mathrm{h}}_{12}$ and $R^{\mathrm{h}}_{23}$ separately as compared to its combination in {\protect\NoHyper\citet{KuruvillaAghanim21}\protect\endNoHyper}.

It is informative to ask how the constraints from $\mathcal{R}$ fares against those obtained from clustering statistics.  In order to answer that question, we compare the constraints obtained in this work with those  obtained from the redshift space halo power spectrum, and the halo bispectrum in \cite{Hahn+20}. Comparing against the constraints from the redshift-space power multipoles for $k_{\mathrm{max}}=0.2\,h\,\mathrm{Mpc}^{-1}$ (which is closest to the $r_{\mathrm{min}}$ considered in this work), $\mathcal{R}$ obtains a factor of improvement of \{2.3, 3.6, 4.5, 5.4, 5.7\} for \{$\Omega_{\mathrm{m}}$, $\Omega_{\mathrm{b}}$, $h$, $n_{\mathrm{s}}$, $M_\nu$\}, respectively. However  it slightly reduces to  \{1.4, 2.9, 3.1, 3.3, 2.5\} when comparing against the constraints from power spectrum when $k_{\mathrm{max}}=0.5\,h\,\mathrm{Mpc}^{-1}$. This improvement over power spectrum (a two-point summary statistics) is not surprising as $\mathcal{R}$ is based on the first moment of the three-point relative velocity statistics. Hence it is interesting to compare the constraints against those obtained from the redshift-space bispectrum monopole, and when using all triangular configurations for $k_{\mathrm{max}}=0.2\,h\,\mathrm{Mpc}^{-1}$, $\mathcal{R}$ still obtains a factor of improvement of \{1.8, 2.9, 3.2, 3.1, 1.8\} for \{$\Omega_{\mathrm{m}}$, $\Omega_{\mathrm{b}}$, $h$, $n_{\mathrm{s}}$, $M_\nu$\}, respectively. But when considering a larger set of triangular configurations for the bispectrum with $k_{\mathrm{max}}=0.5\,h\,\mathrm{Mpc}^{-1}$, there is less constraining power for $\mathcal{R}$ with a factor of  \{0.7, 1.0, 1.0, 0.9, 0.4\} for \{$\Omega_{\mathrm{m}}$, $\Omega_{\mathrm{b}}$, $h$, $n_{\mathrm{s}}$, $M_\nu$\}, respectively. This is not surprising, as the bispectrum monopole in this case, probes further into the nonlinear scales while $\mathcal{R}$ was analysed for triangular configurations with separation scales of $40\,h^{-1}\mathrm{Mpc}$ and above.

As it currently stands one of the limitations to applying the new statistic, $\mathcal{R}(\triangle_{123})$, directly to any observational data is the lack of an estimator to measure the three-point mean radial relative velocities using the LOS velocities (as it will be the LOS velocity which can be measured from either the peculiar velocity surveys or kSZ experiments). In the case of the mean pairwise velocity, \cite{Ferreira+99} has shown how to construct such an estimator. Similarly for the case of $R_{ij}(\triangle_{123})$, and $\mathcal{R}(\triangle_{123})$, we will be constructing such an estimator in a future work.  In the case of the mean pairwise velocities, an alternative estimator exists using each tracers' transverse velocity component \citep{Yasini+19}. On the other hand, the three-point mean relative velocity consists of a non-vanishing mean transverse component in the plane of the triangle (unlike in the case of the pairwise velocity which has its transverse component equal to zero). Thus we would construct an estimator for the non-vanishing three-point mean transverse relative velocity in a future work. And furthermore the analysis we presented here took only the radial component into consideration, and hence a combination of both radial and transverse components of the three-point mean relative velocity could further improve the chances of constraining the cosmological model accurately.

Another caveat which we has not discussed in this work is the mass dependence of the optical depth parameter, which has been shown to increase as the halo mass increases \citep[e.g.][]{Battaglia16}.  We considered it as an averaged quantity [as shown in Eqs.~(\ref{eq:ksz-pair}) and (\ref{eq:ksz-triplet})]. However we do not envision the mass dependence to affect $\mathcal{R}(\triangle_{123})$, as long as the ratio is done using the same mass bin. On the other hand assuming a fixed cosmology, we could consider a scenario where measuring $R^{\mathrm{h}}_{23}$ is fixed to a high halo mass bin while measuring $R^{\mathrm{h}}_{12}$ for various mass bins in Eq.~(\ref{eq:ratio-new-statistics}). Thus, it could lead to potentially measuring the (scaled) mass dependence of optical depth (and degenerate with the bias factor) from direct kSZ experiment directly.

\section{Conclusions}
\label{sec:conclusions}

Determination of neutrino mass using cosmological observables have become one of the main goals for the forthcoming cosmological surveys. However the two-point statistics in general is affected by the $M_\nu$-$\sigma_8$ degeneracy, whether using clustering or relative velocity statistics which limits the potential of constraining neutrino mass from cosmology. With regards to relative velocities, \cite{KuruvillaPorciani20} introduced the three-point mean relative statistics (i.e. the mean relative velocity between pairs in a triplet), and subsequently in \cite{KuruvillaAghanim21} they quantified the cosmological information content in them. It was found to offer substantial information gain when compared to two-point statistics (both power spectrum and mean pairwise velocity), while being competitive with the constraints from the bispectrum.

In this paper, we extended the applications with the mean three-point relative velocity statistics, and introduced a new ratio statistic $\mathcal{R}$ [Eq.~(\ref{eq:ratio-new-statistics})]  which is unaffected by $\sigma_8$. This enables to constrain neutrino mass, in addition to other cosmological parameters, independent of $\sigma_8$. Moreover, in the context of kSZ experiments this statistic is independent of optical depth, hence circumventing the optical depth degeneracy which currently acts a limiting factor in the determination of cosmological parameters from the kSZ experiments. Furthermore, the leading order perturbation theory prediction suggests that $\mathcal{R}$ will be bias independent on linear scales. We verified it by measuring $\mathcal{R}$ for both halos and matter, and found that the bias is consistent with one at 1--2\% for all triangular configurations we probed in this work ($r_{\mathrm{min}}=40\ h^{-1}\mathrm{Mpc}$ and $r_{\mathrm{max}}=120\ h^{-1}\mathrm{Mpc}$).

We also studied the effect of summed neutrino mass on $\mathcal{R}$, and found that as the neutrino mass increases the amplitude of $\mathcal{R}$ decreases. This can be understood by the fact that due to the free streaming of neutrinos the collapse of matter slows down, and by the virtue that $\mathcal{R}$ acts as a proxy to the mean infall velocity between pairs in a triplet, $\mathcal{R}$ decreases as $M_\nu$ increases. 

We used the Fisher-matrix formalism to quantify the information content in $\mathcal{R}$, where the necessary derivatives and the covariance matrices were directly measured from the Quijote suite of simulations. We utilised 15,000 realisations of the reference cosmology to compute the covariance matrix, and  the partial derivatives were also computed directly from the simulations. We find that constraints obtained from $\mathcal{R}$ has a factor of 6.2--12.9 improvement when compared against the constraints obtained from the mean pairwise velocity. When compared against the power spectrum and bispectrum, it still achieves an improvement in the constraints with a factor of 2.3--5.7 and 1.8--3.2, respectively.

In summary we have introduced a new statistic based on the mean radial relative velocity between pairs in a triplet and shown that it can act as robust cosmological observable which could lead to sizeable information gain in comparison to the mean radial pairwise velocity. One of the limitation of the kSZ experiments is the optical depth degeneracy, and breaking this degeneracy requires some form of external data set \citep[for e.g. using fast radio bursts as suggested in][]{Madhavacheril+19}.
This new statistic thus provides a way forward in which the cosmological parameters can be constrained using data from future kinetic Sunyaev--Zeldovich experiments alone, without being affected by the optical depth parameter. 

\begin{acknowledgements}
We would like to thank Nabila Aghanim and Francisco Villaescusa-Navarro for useful discussions. JK acknowledges funding for the ByoPiC project from the European Research Council (ERC) under the European Union's Horizon 2020 research and innovation program grant agreement ERC-2015-AdG 695561. We are thankful to the  community for developing  and  maintaining open-source software packages extensively used in our work, namely \textsc{Cython} \citep{cython}, \textsc{Matplotlib} \citep{matplotlib} and \textsc{Numpy} \citep{numpy}.
\end{acknowledgements}

\setlength{\bibhang}{2.0em}
\setlength\labelwidth{0.0em}
\bibliographystyle{aa}
\bibliography{main}



\end{document}